\documentclass[
    ,final            
  ]
  {aipproc}

\layoutstyle{6x9}
\usepackage{epstopdf}

\begin{document}

\title{Double Longitudinal Spin Asymmetries of Inclusive Charged Pion Production in Polarized $p+p$ Collisions at 200 GeV}


\author{Adam Kocoloski (for the STAR Collaboration)}{
  address={Massachusetts Institute of Technology\linebreak
  77 Massachusetts Avenue,  Cambridge, MA  02139,  USA}
}

\begin{abstract}
A primary goal of the STAR Spin program at RHIC is the measurement of the polarized gluon distribution function $\Delta G$, which can be obtained from a global analysis incorporating measurements of the double spin asymmetry $A_{LL}$ in various final state channels of polarized $p+p$ collisions.  Final states with large production cross sections such as inclusive jet and hadron production are analyzed as the program moves towards the measurement of $A_{LL}$ in the theoretically clean channel of prompt photon production.  The channels $p+p \rightarrow \pi^{+/-} + X$ are unique in that the ordering of the measurements of $A_{LL}$ in these two channels is sensitive to the sign of $\Delta G$.  Moreover, STAR has already established the procedure for the identification of charged pions and the calculation of their production cross-sections over a broad kinematic range.

This contribution will present first measurements of double longitudinal spin asymmetries for inclusive charged pion production extracted from 3 $pb^{-1}$ of data at $\sqrt{s}$=200 GeV and 50\% beam polarizations.  The asymmetries are calculated over the transverse momentum region $2<p_T<10$ GeV/c and compared with theoretical predictions incorporating several gluon polarization scenarios.  A systematic bias introduced by the selection of charged pions from events satisfying electromagnetic energy triggers will be discussed and estimated using Monte Carlo. 
\end{abstract}

\maketitle



The polarized $p+p$ physics program at the Relativistic Heavy Ion Collider (RHIC) offers a unique opportunity to constrain the  gluon contribution to the spin of the proton through the measurement of double longitudinal spin asymmetries in a variety of final states \cite{Bunce:2000uv}.  In the QCD factorization scheme, the double spin asymmetry $A_{LL}$ for a particular final state contains contributions from one or more subprocesses, each of which can be expressed as a convolution of a short-range calculable partonic asymmetry and a set of long-range distribution functions. Final states such as inclusive jet and hadron production contain contributions from gg and qg scattering and are thus sensitive to $\Delta g(x)$ at leading order.  These final states also have large production cross sections and are reasonable first steps for the RHIC Spin program.  In particular, measurements of $A_{LL}$ for inclusive charged pion production are interesting because the ordering of the measurements is sensitive to the sign of the gluon polarization.  In the high $p_T$ region where qg scattering plays the dominant role, the sign of $A_{LL}$ is determined by the product of the quark and gluon polarizations, while production of the two charged pion species is dominated by different quark flavors with opposite polarization signs.


The double spin asymmetry $A_{LL}$ for a particular process can be defined as a ratio of polarized and unpolarized cross sections.  Equivalently, one can write the asymmetry in terms of beam polarizations P, relative luminosities R, and spin-dependent yields N as

\begin{equation}
A_{LL} = \frac{1}{P_1 P_2}\frac{(N^{++} - RN^{+-})}{(N^{++} + RN^{+-})}.
\end{equation}

The charged pion $A_{LL}$ measurements presented in this contribution were obtained using 1.6 $pb^{-1}$ of data collected by the STAR experiment during the 2005 RHIC run \cite{Ackermann:2002ad}.  Events were selected by hardware triggers requiring a coincidence of signals in the Beam Beam Counter (BBC) scintillator tiles at forward rapidity.  The STAR minimum-bias trigger is completely defined by this trigger condition and collected data at a large prescale throughout the run.  The bulk of the statistics are obtained from "jet patch" triggers that also look for deposits of electromagnetic energy in $\Delta \eta \times \Delta \phi$ = $1.0 \times 1.0$ patches of the barrel electromagnetic calorimeter (BEMC).  These triggers were only implemented in the region $0 < \eta < 1$.

\begin{figure}
\begin{minipage}[t]{.4\textwidth}
	\includegraphics[width=.99\textwidth]{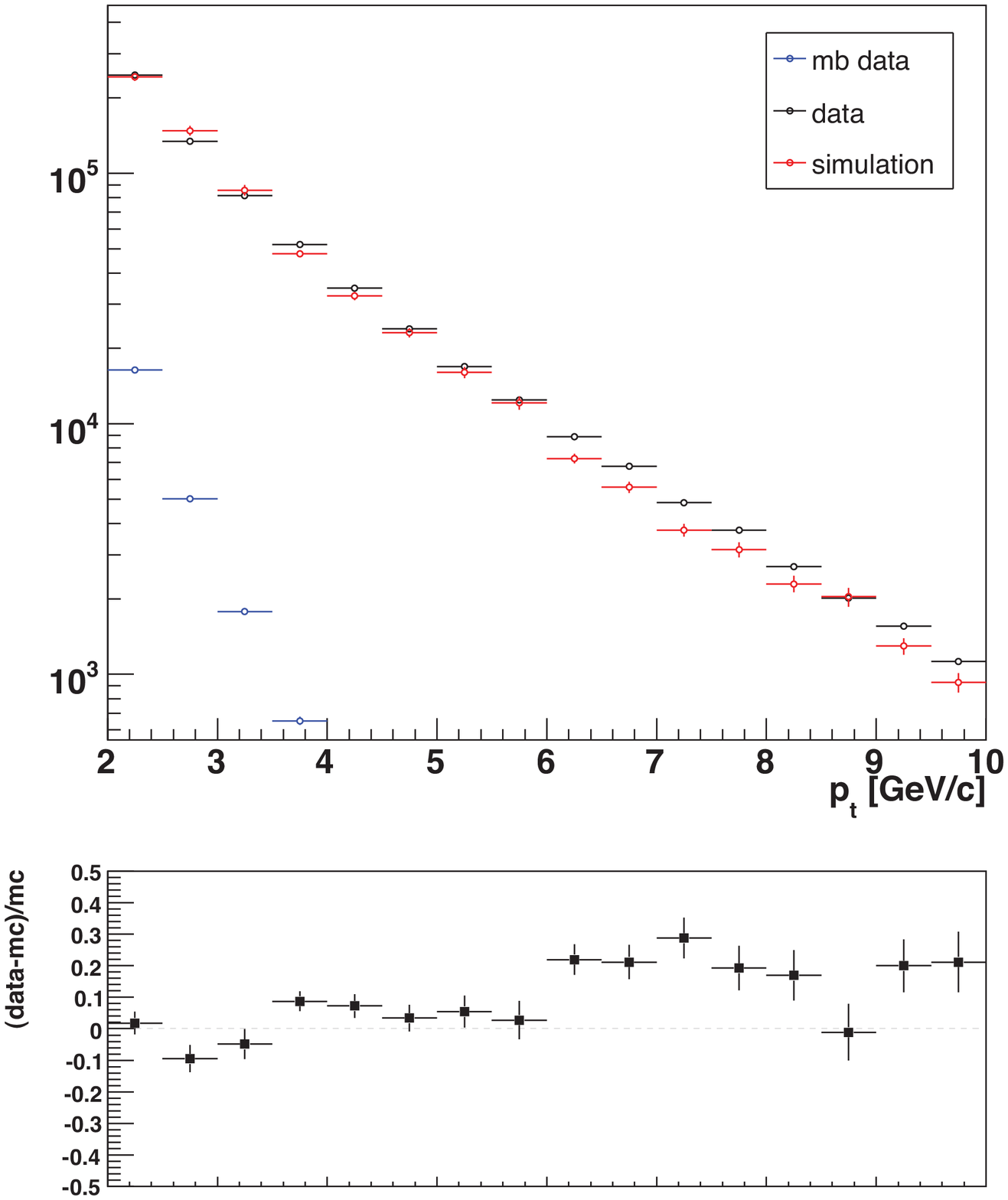}
\end{minipage}
\begin{minipage}[t]{.4\textwidth}
	\includegraphics[width=.99\textwidth]{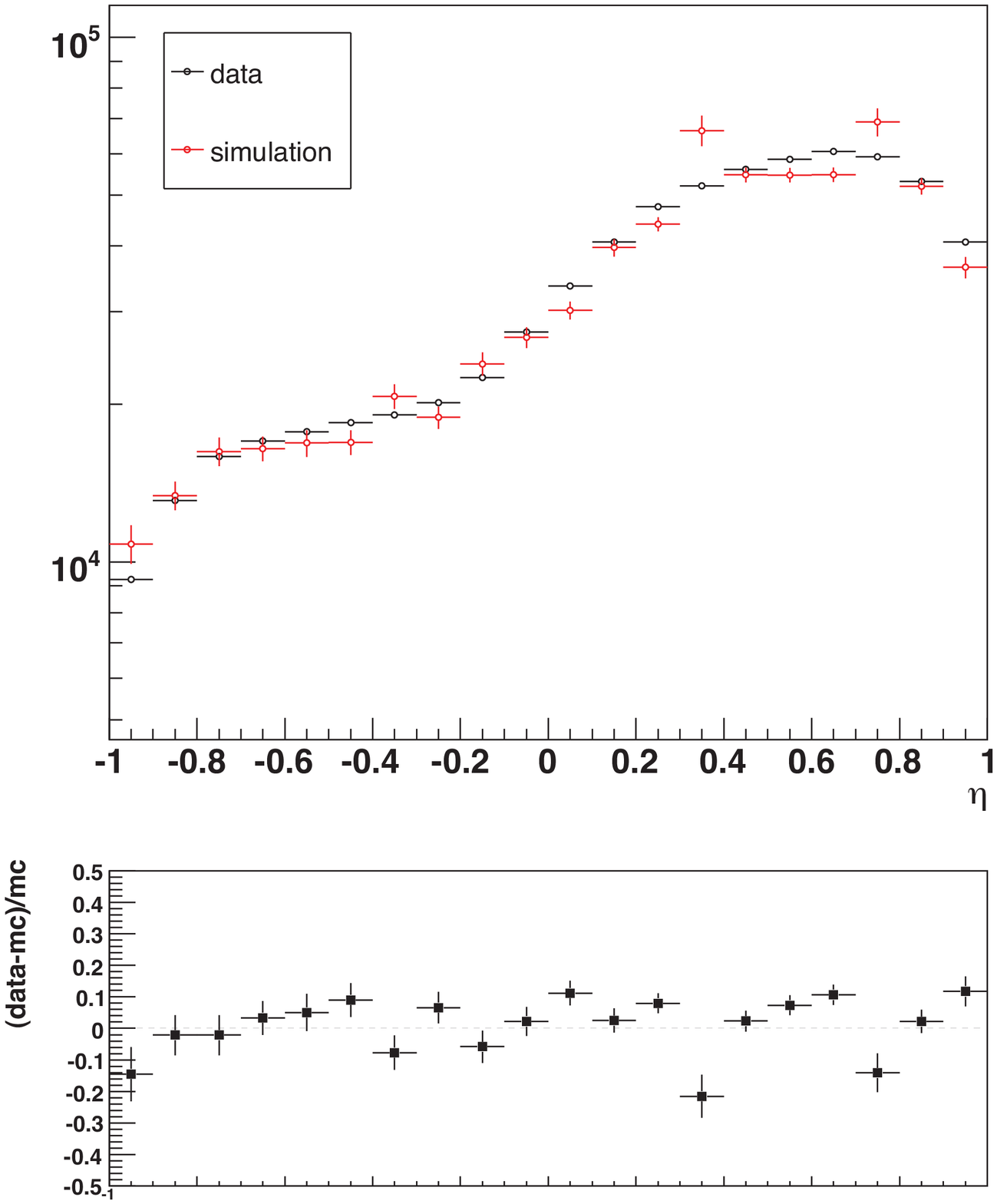}
\end{minipage}
\caption{Distribution of $\pi^{+/-}$ accepted for $A_{LL}$ measurements and comparison to simulation.  The left panel shows the $p_T$ dependence of jet patch triggered $\pi^{-}$ in data and simulation.  The minimum-bias triggered data is overlaid to demonstrate the necessity of using the jet patch data.  The right panel shows the pseudo-rapidity dependence of jet patch triggered $\pi^{+}$, where the effect of the trigger in the positive $\eta$ region is clearly visible.
}
\label{fig:datamc}
\end{figure}

STAR reconstructs and identifies charged pions using a large Time Projection Chamber (TPC) \cite{Anderson:2003ur}.  The TPC is situated inside of a 0.5T magnetic field that allows for the measurement of transverse momenta up to 20 GeV/c.  Particle identification is accomplished via measurements of the energy loss of the TPC hits and is effective at identifying pions over the momentum range $1 < p < 10$ GeV/c.  In this analysis a high-quality sample of charged pions with $2 < p_T < 10$ GeV/c and $-1 < \eta < 1$ was used.  Stringent cuts on the number of TPC hits per track (25) and the distance of closest approach to the primary vertex (1 cm) were also applied.  The purity of the pion sample is estimated to  be greater than 90\%.  The agreement between data and simulation for the accepted charged pion sample is detailed in Figure \ref{fig:datamc} and allows the use of simulations to characterize some of the systematic effects present in the data sample.  This technique for pion reconstruction and identification has already been used to calculate the production cross section for charged pions in minimum-bias $p+p$ collisions \cite{Adams:2006nd}.  The good agreement of that published cross section with NLO pQCD predictions supports the use of QCD factorization in the interpretation of these measurements of charged pion $A_{LL}$.


Measurement of $A_{LL}$ requires independent knowledge of beam polarizations and relative luminosities in addition to the spin-dependent yields.  Polarizations $P_{1,2}$ were measured several times a day using the RHIC CNI polarimeter.  This polarimeter is ultimately calibrated against a hydrogen gas jet polarimeter to set the absolute polarization scale, but the online measurements used in this analysis are only known to an uncertainty of $\sim20\%$.  The luminosities per bunch crossing are obtained using the BBC and are known to $\delta R/R$ = $10^{-3}$.  Typical values for the beam polarizations and relative luminosities were $P_{1,2} = 50\%$ and R = 1.1.

\begin{figure}
\begin{minipage}[t]{.5\textwidth}
	\includegraphics[width=.99\textwidth]{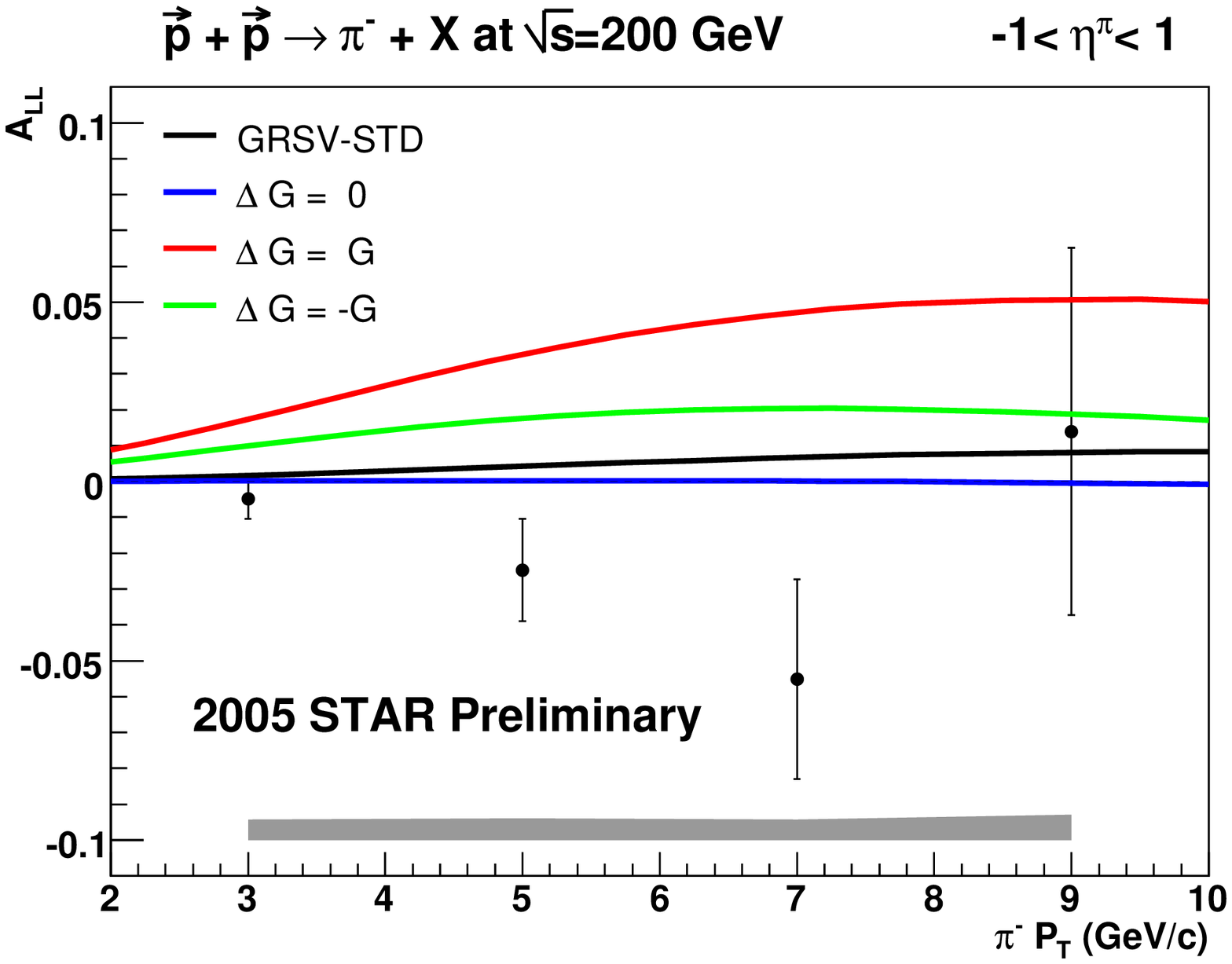}
	\label{fig:all_plus}
\end{minipage}
\begin{minipage}[t]{.5\textwidth}
	\includegraphics[width=.99\textwidth]{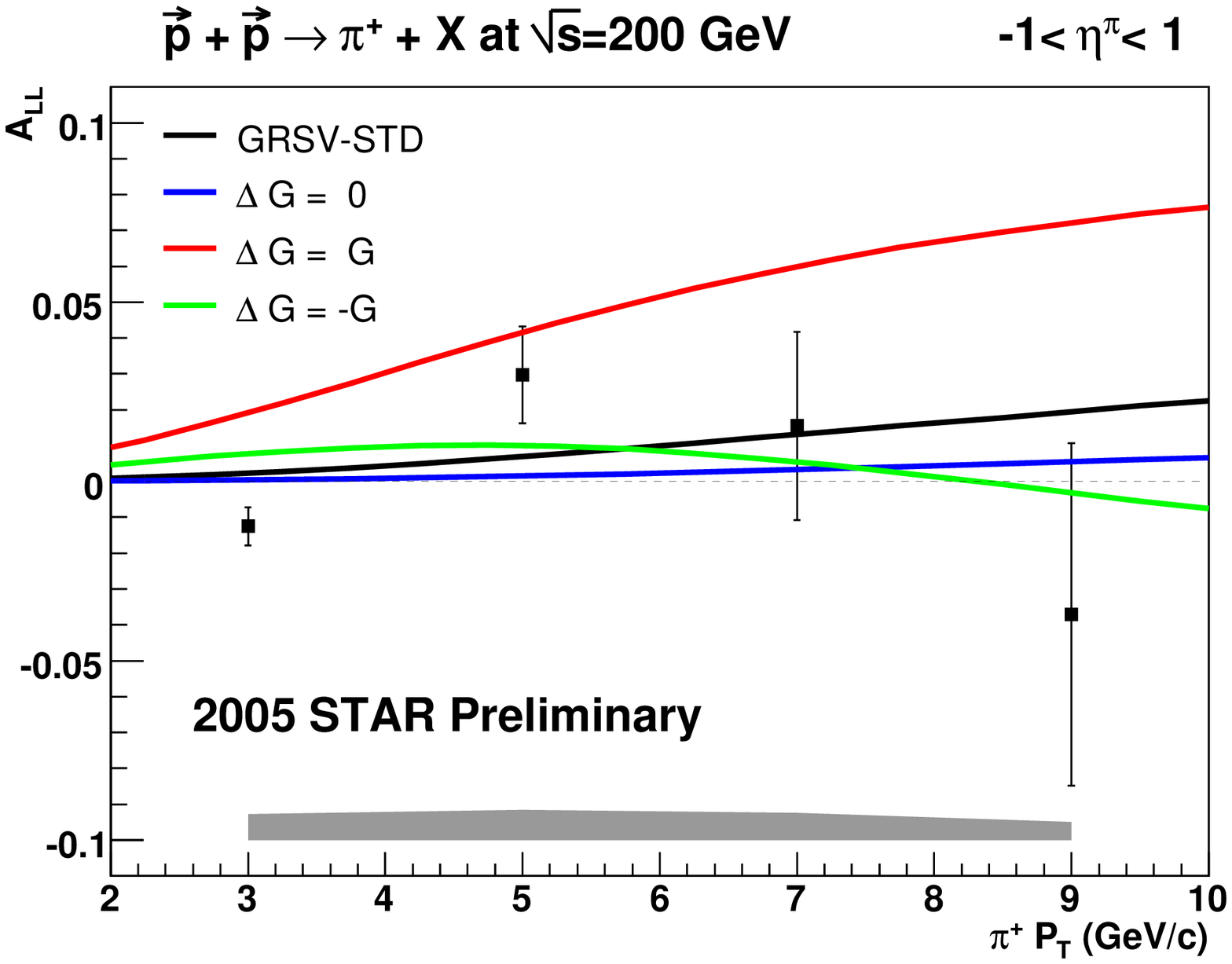}
	\label{fig:all_minus}
\end{minipage}
\caption{Double longitudinal spin asymmetries for inclusive charged pion production.  $A_{LL}(\pi^{-})$ is displayed in the left panel and $A_{LL}(\pi^{+})$ is on the right.  The asymmetries are compared to theoretical predictions of $A_{LL}$ incorporating various scenarios for the gluon polarization.  The error bars are statistical; point-to-point systematic uncertainties are added in quadrature and shown as the gray band at the bottom of each figure.  A preliminary scale uncertainty of $\sim$40\%  from the uncertainty on the beam polarization measurements is not included.}
\label{fig:all}
\end{figure}

Figure \ref{fig:all} displays STAR's preliminary results for double longitudinal spin asymmetries of inclusive charged pion production in the 2005 run.  The theoretical predictions for $A_{LL}$ are calculated at NLO and use a variety of possibilities for the polarized gluon distribution functions as inputs \cite{Jager:2004jh}.  The black lines are obtained from an input distribution corresponding to the best fit to DIS data on $\Delta G$.  The fragmentation functions for $\pi^{+}$ and $\pi^{-}$ are based on the KKP fragmentation functions \cite{Kniehl:2000hk}.  These fragmentation functions are not charge-separated, so an ansatz was employed that multiplied the favored fragmentation functions by ($1+z$) and the unfavored ones by ($1-z$).

A variety of tests for systematic effects were performed to check the stability of the result.  In particular, the leading systematic uncertainty accounts for the bias introduced by the choice of event triggers for the analysis.  The jet patch triggers that dominate the event sample are biased towards jets with large neutral energy fractions, and as a result the charged pions in the jet that triggered each event  carry an artificially low fraction of the momentum of the outgoing quark.  The effect of these triggers on the measured asymmetries was estimated in simulations to be approximately $5\times10^{-3}$, which is comparable to the statistical error in the first $p_T$ bin.

\begin{figure}
\includegraphics[width=.5\textwidth]{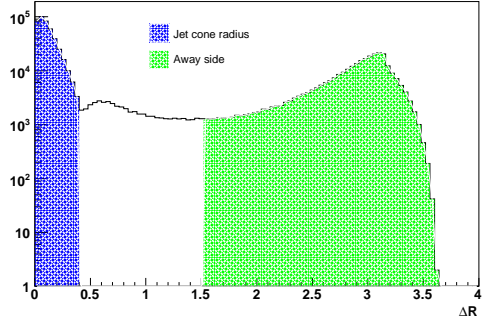}
\caption{Distribution of charged pions relative to the trigger jet as a function of $\Delta R$}
\label{fig:deltaR}
\end{figure}

Figure \ref{fig:deltaR} shows the distribution of charged pions as a function of $\Delta R = \sqrt{(\Delta \eta)^{2} + (\Delta \phi)^{2}}$ relative to the jet that satisfied the trigger requirement.  The blue shaded region indicates pions that fall within the cone radius used for jet reconstruction in STAR.  Pions falling in the green shaded region are termed "away-side" pions.  Measurements of $A_{LL}$ for these two classes of particles are consistent within errors, supporting the calculation of an inclusive $A_{LL}$.  However, the use of the jet patch triggers only introduces a significant fragmentation bias in the near-side sample.  As the spin physics program at RHIC matures and collects additional luminosity,  exclusive measurements restricted to charged pions opposite a trigger jet may well be the more sensitive charged pion $A_{LL}$ measurements.

This contribution presented STAR's first measurements of $A_{LL}$ for inclusive charged pion production in polarized proton collisions at $\sqrt{s}$ = 200 GeV.  The measured asymmetries are compared to a range of predictions for $A_{LL}$ and disfavor the maximal gluon polarization scenario.  Future measurements will benefit from a significant increase in figure of merit as well as the possibility to leverage STAR jet reconstruction to select charged pions opposite a triggered jet.



\bibliographystyle{aipproc}   

\bibliography{AKocoloskiSpin06}

\end{document}